\documentclass[10pt,conference]{IEEEtran}
\IEEEoverridecommandlockouts
\usepackage{cite}
\usepackage{amsmath,amssymb,amsfonts}
\usepackage{algorithmic}
\usepackage{graphicx}
\usepackage{textcomp}
\usepackage{xcolor}
\def\BibTeX{{\rm B\kern-.05em{\sc i\kern-.025em b}\kern-.08em
    T\kern-.1667em\lower.7ex\hbox{E}\kern-.125emX}}

\usepackage{float}
\usepackage{placeins}
\usepackage{url}
\usepackage{amsmath}
\usepackage{multirow}
\usepackage{subcaption}

\begin{document}

\title{Designing a Bot for Efficient Distribution of Service Requests}

\author{
    \IEEEauthorblockN{Arkadip Basu, Kunal Banerjee}
	
	\IEEEauthorblockA{Walmart Global Tech, Bangalore, India\\
	\{Arkadip.Basu, Kunal.Banerjee1\}@walmartlabs.com
	}
}

\maketitle

\begin{abstract}
The tracking and timely resolution of service requests is one of the major challenges in agile project management.
Having an efficient solution to this problem is a key requirement for Walmart to facilitate seamless collaboration across its different business units.
The Jira software is one of the popular choices in industries for monitoring such service requests.
A service request once logged into the system by a reporter is referred to as a (Jira) ticket which is assigned to an engineer for servicing.
In this work, we explore how the tickets which may arise in any of the Walmart stores and offices distributed over several countries can be assigned to engineers efficiently.
Specifically, we will discuss how the introduction of a bot for automated ticket assignment has helped in reducing the disparity in ticket assignment to engineers by human managers and also decreased the average ticket resolution time -- thereby improving the experience for both the reporters and the engineers.
Additionally, the bot sends reminders and status updates over different business communication platforms for timely tracking of tickets; it can be suitably modified to provision for human intervention in case of special needs by some teams.
The current study conducted over data collected from various teams within Walmart shows the efficacy of our bot.
\end{abstract}

\begin{IEEEkeywords}
Project Management, Jira Software, Enterprise Bot, Resolution Time
\end{IEEEkeywords}

\section{Introduction}\label{sec:intro}
Walmart runs its retail industry business on multiple online e-commerce portals as well as in the traditional brick and mortar stores and clubs, which are spread across 27 countries.
The overall business is supported by collaborative efforts from different teams, which are also spread across different countries.
Jira~\cite{jira}, developed by Atlassian, is an enterprise grade tool to raise and track the operational activities of an industry.
An associate (reporter) raises a service request (ticket) on the Jira portal, which is then resolved by another engineer.
Typically, service-level agreements (SLAs) are made to guarantee the level of service an associate may expect along with a stipulated time period within which a service request is supposed to be resolved.
Therefore, within the deadline mentioned in the SLA, the engineer tries to resolve the issue and sends the solution for review; once the solution is found to be satisfactory, the ticket is marked as closed.
During the process, the reporter and the engineer may communicate with each other to understand the issue and then resolve it appropriately.

Previously, to assign work on a new ticket, engineers had to visit the service board in frequent intervals; the process also at times involved members from technology project management office (Tech-PMO) and the scrum master. 
If the process got stuck, the tickets were put on blocked state for long periods.
Eventually, people forgot to proceed on the tickets which further delayed the resolution.

Now, after the introduction of the bot, human interference can be eliminated (although some teams still prefer to keep a human manager in the loop); the bot scans the Jira portal at periodic intervals, finds the tickets which are new and unassigned and tags these to the available engineers.
If the SLA is about to be breached, the bot notifies the team across multiple business platforms, namely, Slack, Microsoft Teams and Outlook.
Moreover, if a ticket gets blocked for any reason and stays pending, the bot notifies both the reporter and the assigned engineer to resolve the issue.

Walmart being a global brand has footprint in several countries across different continents.
This work, however, is based on data collected from reporters and engineers residing in the countries: Brazil, Canada, Chile, China, India, Japan, Mexico, South Africa and USA, with majority of the engineers located in USA and India.
Everyday thousands of service requests are submitted in Walmart across different boards.
The count varies time to time based on business events and typically peaks during the holiday seasons.
If we look at a granular level, in respective board (department) the count is about 30 tickets per business day.
Our bot is deployed on 11 different boards as of now.
In this work, we will discuss how the automated ticket assignment done by our bot reduced the disparity in ticket assignment to engineers which was earlier done by human managers, and also decreased the average ticket resolution time -- thereby, improving the experience for both the reporters and the engineers.
Furthermore, the bot sends reminders and status updates over different business communication platforms for timely tracking of tickets.
We shall also discuss ideas that worked in practice and others that did not.
It is important to note that the methodology adapted in this work is applicable to other project management softwares as well and not just to the Jira software.

The paper is organized as follows.
Section~\ref{sec:workflow} briefly describes the workflow of a Jira ticket.
Section~\ref{sec:bot} covers the bot deployment.
Section~\ref{sec:result} presents our experimental results related to efficient distribution of the tickets and how it translated to reduction in average ticket resolution time.
It additionally includes the gist of our survey conducted with the Walmart engineers about how satisfied they are with the bot.
Some related work are covered in Section~\ref{sec:literature}.
Section~\ref{sec:concl} concludes the paper.

\section{Jira Ticket Workflow}\label{sec:workflow}
\begin{figure}
\centering
\includegraphics[width=\columnwidth]{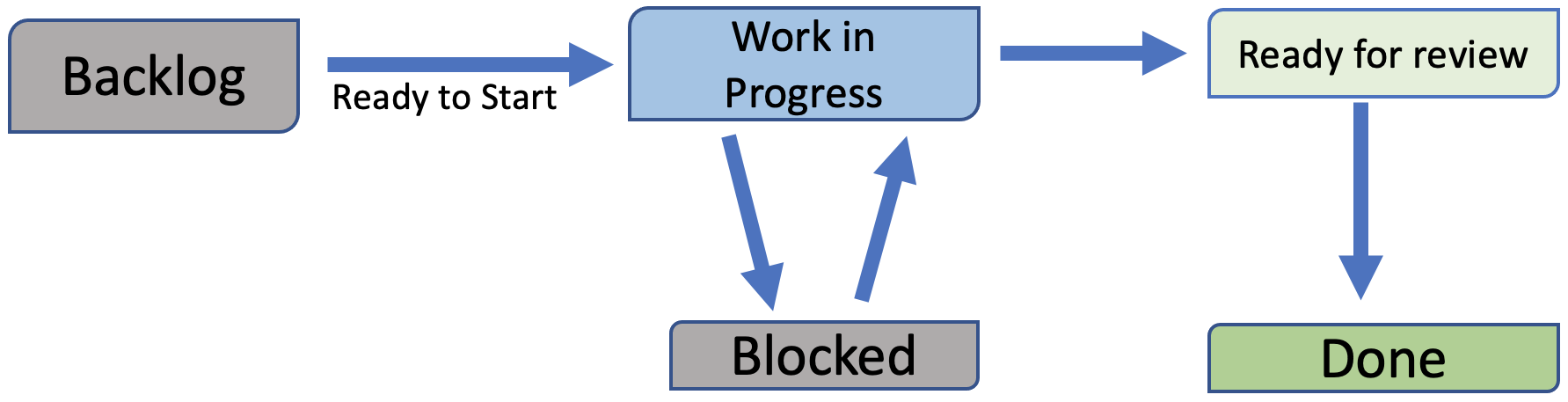}
\caption{A typical workflow of a ticket in the Jira software.}\label{fig:workflow}
\end{figure}

The typical workflow of a Jira ticket is shown in Fig.~\ref{fig:workflow} and is also described below.
\begin{enumerate}
\item \textbf{Backlog:} When a reporter raises a ticket mentioning a business requirement, this is the initial state allotted to the ticket which indicates that it is due for attention from the engineer.
\item \textbf{Ready to start:} This is an optional status which is applied when the assigned engineer indicates that he is ready to work on the ticket.
\item \textbf{Work in progress:} This status indicates that the engineer is working on the ticket and it is expected that the resolution will be provided within the SLA deadline.
\item \textbf{Blocked:} If during the work there is a lack of information or the engineer requires assistance to process the issue, then the ticket is put on blocked state -- this draws attention of the Tech-PMO and the reporter. Post consultation the ticket is again restored to ``Work in progress.''
\item \textbf{Ready for review:} When the engineer confirms that (s)he is done from his/her end, then the ticket is transitioned to this state. This is a request to the reporter to validate the resolution.
\item \textbf{Done:} Post validation from the reporter (and other stakeholders, if needed) or after all the checks are met, the ticket is marked as ``Done'' (also known as ``Resolved'') -- it ends the journey of the ticket's workflow.
\end{enumerate}
Please note that rarely a ticket may be re-opened in case an issue is not properly fixed or it resurfaces in a different module; in such a case, the status of the ticket may be changed from ``Done'' to ``Backlog'' or more commonly to ``Assigned'' when it is re-assigned to the same engineer who had earlier worked on it.
Moreover, some tickets do not undergo the whole workflow and reach the ``Done'' state - this typically happens in case of duplicate tickets or when the engineer and the reporter reach a mutual solution informally in case of low priority tickets.

\section{Bot Deployment}\label{sec:bot}
\begin{figure}
\centering
\includegraphics[width=0.9\columnwidth]{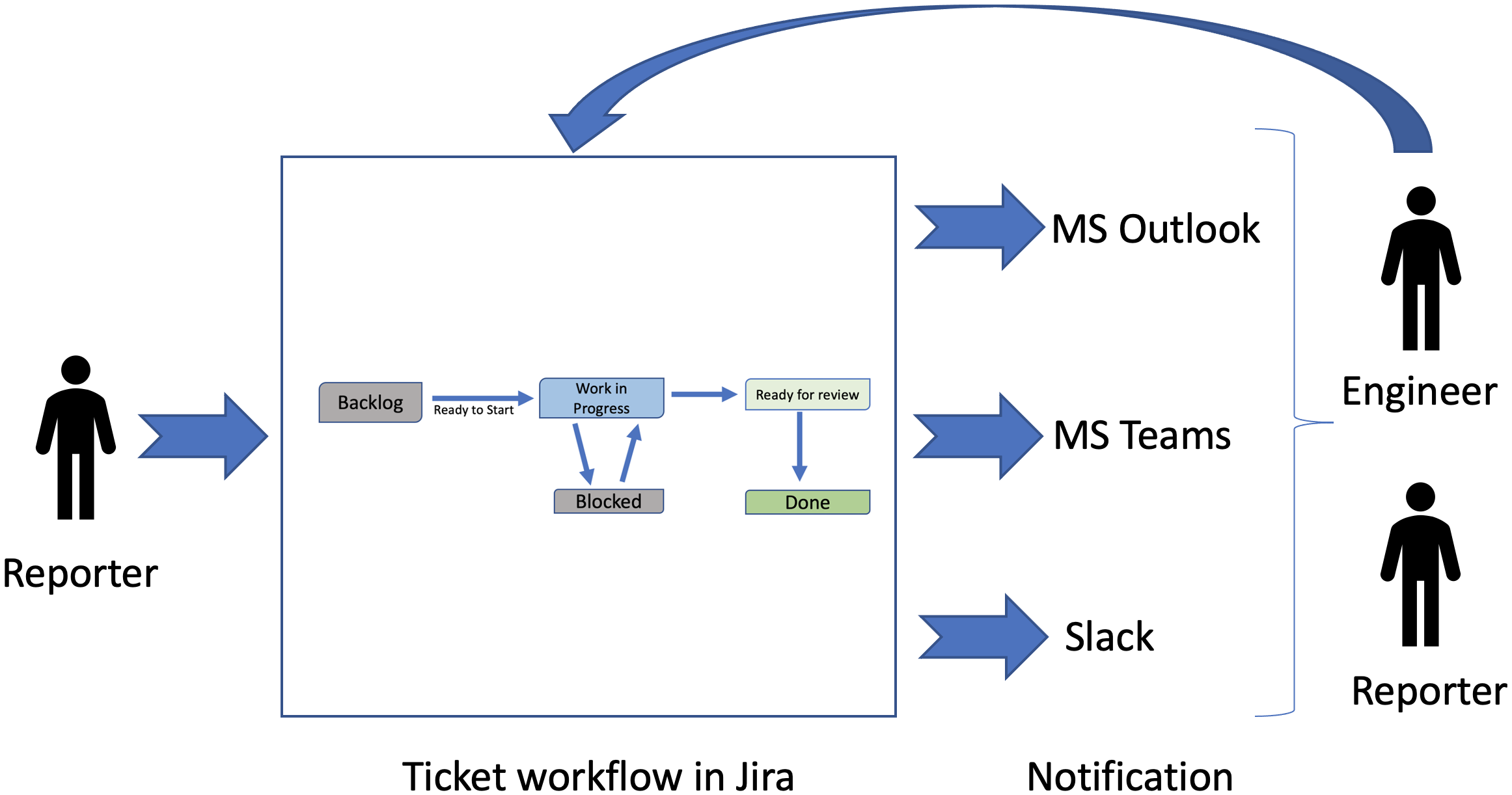}
\caption{Bot deployment.}
\label{fig:bot_deployment}
\end{figure}

Our bot is currently deployed for 11 boards assisting teams working on networking, application delivery, telecommunication, information technology and many more.
The pool of engineers in each team is provided to the bot which can be edited to account for joining and separation of employees, or leaves applied for by them.
The bot assigns the tickets to the available set of engineers in a round-robin manner, and therefore a ticket stays in an unassigned state for a negligible amount of time (which sometimes used to take a few days prior to  bot deployment).
The bot tracks the progress of the ticket throughout its workflow; if it stays in the same state (other than Done) above a threshold amount of time as suggested by the respective team, then it sends reminders to both the engineer and the reporter across different business communication platforms.
Currently, our bot is configured to communicate across Slack, Microsoft Teams and Outlook -- a team may choose to get notified through some or all of these platforms.
In addition, there is a communication channel (Slack or MS Teams) dedicated to each team where the bot populates the information about each ticket assignment to a teammate along with updates about the tickets' states -- this is primarily done for review purpose.

\section{Experimental Results}\label{sec:result}
In this section, we report the results based on the data collected from 6 teams out of 11 who have adopted our bot.
These teams are named as Team1 to Team6 to maintain their anonymity.
We leave out 5 teams because they have made distinctive customization to our bot (with some of them embracing a ticket assignment policy that involves both automated assignment by the bot and manual assignment by a manager) which makes analysis in a unified manner infeasible.

\subsection{Distribution of the tickets}
\begin{table*}
\centering
\caption{Comparison of ticket distribution pre and post bot deployment.}\label{tab:pre-post}
\begin{tabular}{|l|r|r|r|r|r|r||r|r|r|r|r|r|}
\hline
Team &  \multicolumn{6}{c||}{Pre bot} & \multicolumn{6}{c|}{Post bot}\\
\hline
     & \#tickets & \#engg   & median & max & avg   & std   & \#tickets & \#engg & median & max & avg   & std \\
\hline
Team1  & 931       & 14     & 41.50  & 272 & 66.50 & 78.34 & 466       & 12     & 15.50  & 175 & 38.83 & 49.37\\
Team2  & 672       & 18     & 31.00  & 150 & 37.33 & 35.85 & 377       & 18     & 17.50  &  69 & 20.94 & 16.05\\
Team3  & 673       & 14     & 24.50  & 335 & 48.07 & 83.10 & 549       & 10     & 28.00  & 128 & 54.90 & 53.62\\
Team4  &  89       & 14     &  3.00  &  19 &  6.36 &  6.66 &  62       &  9     &  5.00  &  18 &  6.89 &  5.38\\
Team5  & 192       &  9     & 16.00  &  61 & 21.33 & 18.92 &  68       &  9     &  3.00  &  28 &  7.56 &  9.48\\
Team6  & 262       &  4     & 65.50  &  83 & 65.50 & 19.78 & 250       &  6     & 49.50  &  61 & 41.67 & 12.68\\
\hline
\end{tabular}
\end{table*}

\begin{figure*}
 \centering
 \begin{subfigure}[b]{0.48\textwidth}
  \centering
  \includegraphics[width=\textwidth]{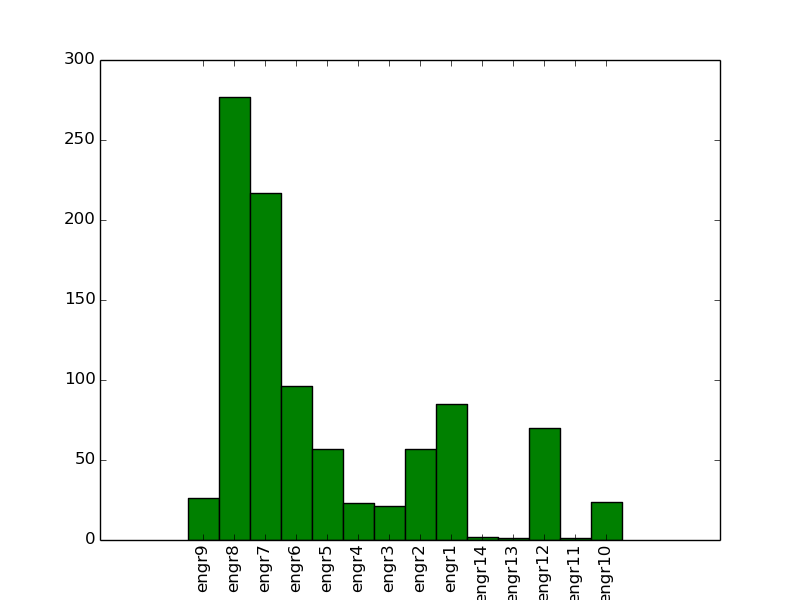}
  \caption{Pre bot deployment}
  \label{fig:prebot_ddi_tickets}
 \end{subfigure}
 \hfill
  \begin{subfigure}[b]{0.48\textwidth}
  \centering
  \includegraphics[width=\textwidth]{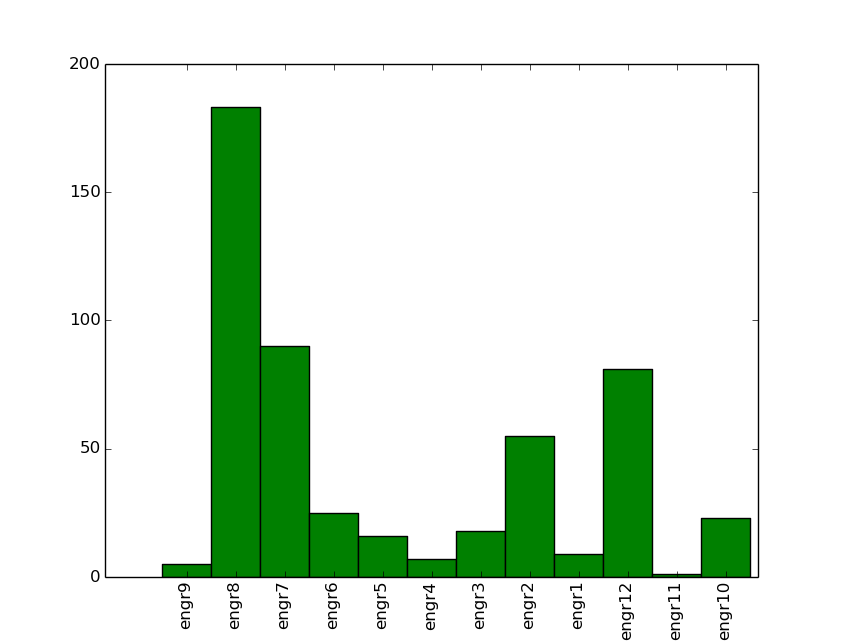}
  \caption{Post bot deployment}
  \label{fig:postbot_ddi_tickets}
 \end{subfigure}
 \caption{Number of tickets assigned to an engineer in Team 1.}
 \label{fig:ddi_tickets}
\end{figure*}

Traditionally, a manager used to manually assign the tickets to the engineers in the team.
This often resulted in a skewed distribution which sometimes led to dissatisfaction among some of the engineers.
Presently, the bot assigns the submitted tickets to the engineers in a team evenly in a round-robin fashion.
We have tried different algorithms for the ticket assignment problem; however, none of these produced a better result -- some of these approaches are discussed subsequently in Section~\ref{sec:impractical}.
Fig.~\ref{fig:prebot_ddi_tickets} shows the distribution of the tickets prior to bot deployment for Team1, whereas Fig.~\ref{fig:postbot_ddi_tickets} shows the distribution post bot deployment.
Note that engr13 and engr14 are present in Fig.~\ref{fig:prebot_ddi_tickets} but not in Fig.~\ref{fig:postbot_ddi_tickets} because they had left the team in the interim period.
One may expect that all the tickets assigned to different individuals in Fig.~\ref{fig:postbot_ddi_tickets} should be identical and indeed it would have been so if we had plotted how the tickets are \textit{initially} assigned by the bot (which is equal to the \textit{average} number mentioned in Table~\ref{tab:pre-post} for Team1 under post bot column); however, here we plot the numbers after the engineers internally reassign the tickets among themselves based on their preference, availability, etc.
We observed that such reassignments occur infrequently to moderately among all teams.
An interesting point to note is that Fig.~\ref{fig:prebot_ddi_tickets} and Fig.~\ref{fig:postbot_ddi_tickets} look similar because the engineer (engr8) who resolved the maximum number of tickets prior to bot deployment continues to hold the top spot even after bot deployment, and we see a similar trend with other engineers as well.
This probably underlines the fact that bot deployment does not affect the relative performance among the engineers.
Table~\ref{tab:pre-post} reports the total number of tickets \textit{resolved} (as considered in this study), the number of engineers in the respective teams, median, maximum and average number of tickets assigned, and the standard deviation in the number of tickets resolved for pre and post bot deployment time period.
The standard deviation captures the skewness in the ticket distribution among the engineers; for all teams, we see that the standard deviation is less for post bot deployment, thus underlining the significance of the bot in achieving a more even distribution for the ticket assignment problem.

\subsection{Average ticket resolution time}
\begin{table}
\centering
\caption{Comparison of average ticket resolution time pre and post bot deployment.}\label{tab:time}
\begin{tabular}{|l|r|r|}
\hline
Team  & Pre bot & Post bot\\
\hline
Team1 & 11d:21h &  7d:02h\\
Team2 & 51d:23h & 13d:06h\\
Team3 &  8d:23h &  7d:19h\\
Team4 & 53d:02h & 45d:20h\\
Team5 & 58d:23h & 25d:13h\\
Team6 & 61d:15h & 21d:19h\\
\hline
\end{tabular}
\end{table}

\begin{figure*}
 \centering
 \begin{subfigure}[b]{0.48\textwidth}
  \centering
  \includegraphics[width=\textwidth]{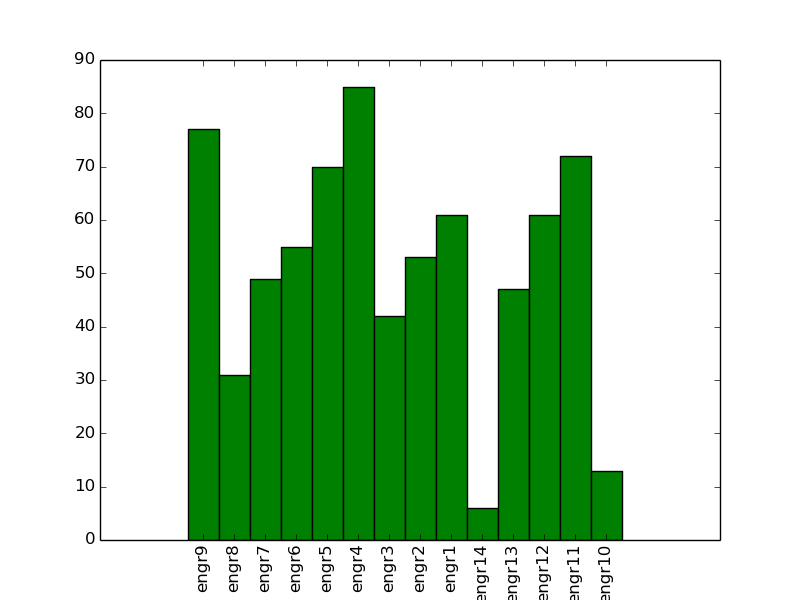}
  \caption{Pre bot deployment}
  \label{fig:prebot_ddi_time}
 \end{subfigure}
 \hfill
  \begin{subfigure}[b]{0.48\textwidth}
  \centering
  \includegraphics[width=\textwidth]{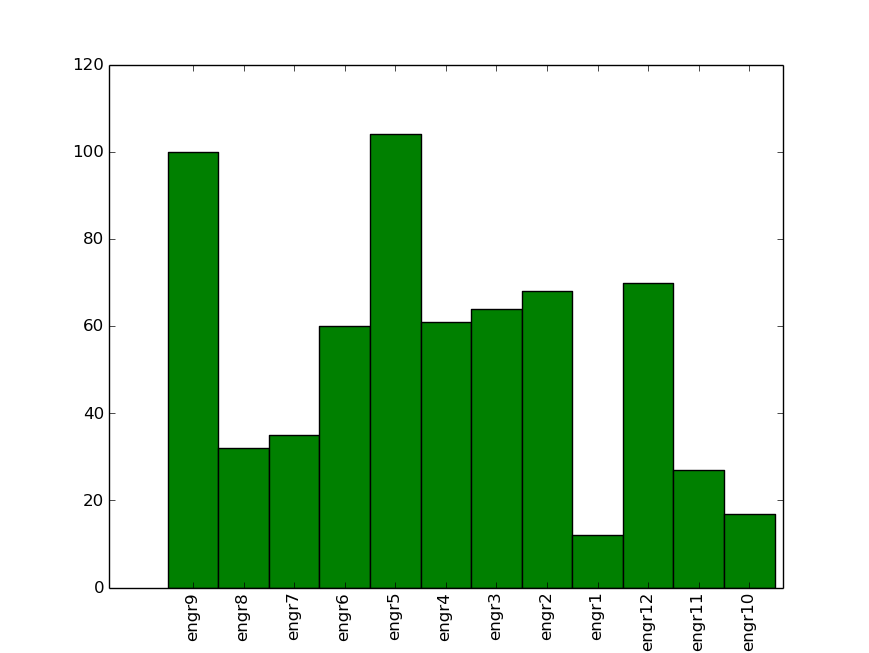}
  \caption{Post bot deployment}
  \label{fig:postbot_ddi_time}
 \end{subfigure}
 \caption{Average number of days taken to resolve a ticket by an engineer in Team 1.}
 \label{fig:ddi_time}
\end{figure*}

In this subsection, we investigate how trying to achieve a balanced distribution of the tickets by our bot additionally affects the average ticket resolution time.
Fig.~\ref{fig:prebot_ddi_time} and Fig.~\ref{fig:postbot_ddi_time} show the average number of days taken by individual engineers in Team1 to resolve a ticket.
As expected, the number of tickets resolved by an engineer and the average time taken to resolve a ticket by that engineer are inversely related.
Table~\ref{tab:time} lists the average ticket resolution time (in terms of days and hours) pre and post bot deployment.
As can be seen from this table, the time is reduced for all teams post bot deployment -- thus improving the experience for both the engineers and the reporters along with positively impacting Walmart business.

\subsection{Engineer satisfaction survey}
\begin{table}
\centering
\caption{Engineer satisfaction survey for the bot.}\label{tab:survey}
\begin{tabular}{|p{3cm}|r|r|r|}
\hline
Question  & Very much & Somewhat & Not at all\\
\hline
Has the bot streamlined the ticket allocation process? & 69\% & 31\% & 0\%\\
\hline
How is the bot helping to maintain SLA? & 46\% & 50\% & 4\%\\
\hline
Does the reminder functionality help to follow up on blocked tickets? & 81\% & 19\% & 0\%\\
\hline
Has the operation process been improved after deployment of the bot? & \multicolumn{2}{|c|}{Yes: 92\%} & No: 8\%\\
\hline
\end{tabular}
\end{table}

We have recently conducted a survey of how satisfied the engineers are with our bot.
From previous experience, we have seen that employees are often reluctant to fill extensive surveys.
Hence, to ensure maximum participation, we had restricted the survey to four questions where the first three questions had three options (Very much / Somewhat / Not at all) and the last one had two (Yes / No).
There was an additional comments section for the participants to share their current experience and recommend future enhancements for the bot.
The engineers from all the six teams had participated.
Almost half of the engineers who are members of these teams and have been working in Walmart during pre and post bot deployment have taken the survey.
We list the questions and the responses in percentage in Table~\ref{tab:survey}.
As can be seen from the table, the responses have been hugely positive although there is some scope of improvement regarding SLA.
One of the most common requests that we received was to integrate our bot with the engineers' calendar to automatically track their leaves, regional holidays, etc.
An example of a negative feedback was that the bot sometimes compounded the panic that an engineer feels when he is already behind in his schedule although similar reminder features have been praised by others.

A similar survey with the reporters would also have been helpful but we found it difficult to find reporters who had faced similar issues both pre and post bot deployment because once the reporter knows the solution, often she can resolve it herself the next time she faces it.
Moreover, the feedback from the reporter is susceptible to the severity of the problem that (s)he may have faced which may have no relation with the performance of the bot.

\subsection{Ideas that did not work in practice}\label{sec:impractical}
\textbf{Expertise-based allocation:} We have often received the request to incorporate an engineer's field of expertise while assigning tickets as prescribed in earlier literature~\cite{wcre13,dretom,bugtossing}.
However, this leads to the problem that the newcomers don't learn new techniques in the process.
Furthermore, later in the event of an expert engineer leaving the organization, we may be left with less competent engineers.

\textbf{Open ticket count-based allocation:} Another request was to keep count of pending tickets for individual engineers so that no one has too many open tickets. However, in this approach, some engineers deliberately kept tickets open so as not to get new tickets assigned to them.

\section{Related Work}\label{sec:literature}
Deploying bots for aiding engineers is a common practice~\cite{botDevops,refbot,repairnator}.
Managing Jira through business platforms such as, Slack, has also been targeted before~\cite{jirio,stratejos}.
There are also bots which process information from Jira boards to manage budgets, sprints and estimates~\cite{stratejos}.
However, to the best of our knowledge, no bot has been designed earlier that is interfaced with the Jira software to assign jobs to engineers at an industry scale; our bot further communicates across multiple business platforms for timely reminders and job tracking.

\section{Conclusion}\label{sec:concl}
In this work, we present how the deployment of a bot for automated assignment of Jira tickets has contributed in achieving a balanced distribution of service requests among the engineers and reduction in ticket resolution time.
The experimental results and the survey conducted underline the efficacy of our design.
In future, we aim to deploy our bot to more teams, learn from their feedback and enhance our bot.
We also intend to find a suitable way to represent the data from the teams which were not covered here to assess the bot's contribution for them.


\bibliographystyle{IEEEtran}
\bibliography{references}

\end{document}